\documentclass[a4paper]{jpconf}
\usepackage{graphicx,citesort,booktabs}
\graphicspath{{./figs/}}
\begin{document}
\title{Machine learning and parallelism in the reconstruction of LHCb
and its upgrade}
\author{Marian Stahl on behalf of the LHCb collaboration}
\address{Physikalisches Institut der Universit\"{a}t Heidelberg, Germany}
\ead{marian.stahl@cern.ch}

\begin{abstract}
After a highly successful first data taking period at the LHC, the LHCb experiment developed a new trigger
strategy with a real-time reconstruction, alignment and calibration for Run~II. This strategy relies
on offline-like track reconstruction in the high level trigger, making a separate offline event
reconstruction unnecessary. To enable such reconstruction, and additionally keeping up with a higher
event rate due to the accelerator upgrade, the time used by the track reconstruction had to be
decreased. Timing improvements have in parts been achieved by utilizing parallel computing
techniques that will be described in this document by considering two example applications. Despite
decreasing computing time, the reconstruction quality in terms of reconstruction efficiency and fake
rate could be improved at several places. Two applications of fast machine learning techniques are
highlighted, refining track candidate selection at the early stages of the reconstruction.
\end{abstract}

\section{Introduction}
The LHCb experiment developed a new trigger strategy with a real-time reconstruction, alignment and
calibration for the Run~II data taking period (2015-2018). Hence, the trigger output is used
to perform physics analyses without the need of an offline event
reconstruction~\cite{LHCb-DP-2016-001}. The main challenge has been to provide a faster and highly
efficient reconstruction with a low rate of "fake" tracks, i.e. charged tracks that do not
correspond to a real particle which passed through the detector. This challenge has been met by
employing parallel computing techniques in timing critical, parallelizable stages, as well as
machine learning in the early selection stages of the reconstruction algorithms. Further
improvements and optimization of the reconstruction software will be essential for the LHCb upgrade
for Run~III data taking (scheduled 2021) where the experiment will move to a trigger-less readout
system and a full software trigger~\cite{LHCb-TDR-012,LHCb-TDR-013,LHCb-TDR-014,LHCb-TDR-015,LHCb-TDR-016}.\\
This document is organized as follows: section~\ref{sec:det} briefly describes the LHCb detector,
the track reconstruction algorithms in place and plans for the upgrade track reconstruction. The use
of parallelism in two timing critical parts of the reconstruction is discussed in
section~\ref{sec:par}. Optimization of track candidate selection using machine learning is subject
to section~\ref{sec:ml}, where two applications will be reviewed.
\newpage
\section{The LHCb detector}\label{sec:det}
\begin{figure}[!htbp]\centering
  \includegraphics[width=0.8\textwidth]{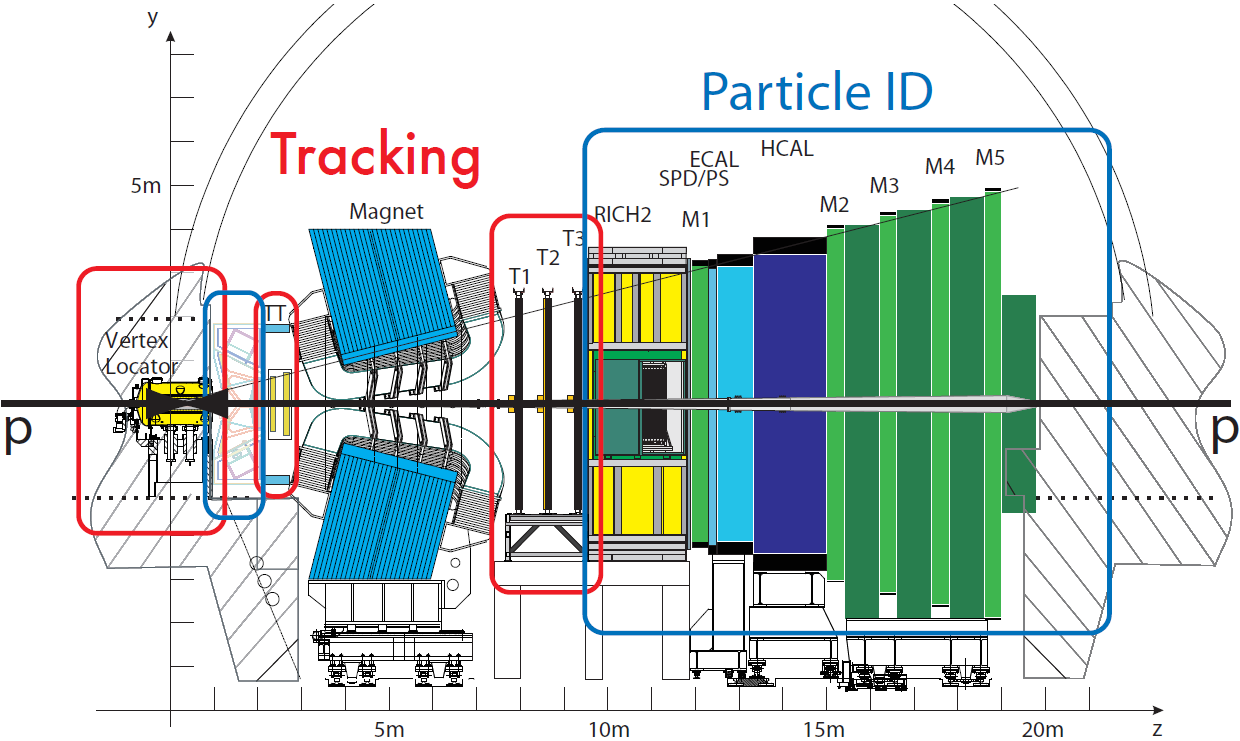}
  \caption{Schematic view of the LHCb detector at the LHC. The tracking subdetectors (VELO, TT,
  T stations) have been highlighted in red and the particle identification subdetectors (RICH1/2,
  calorimeters and muon system) in blue.}
  \label{fig:detector}
\end{figure}

The LHCb detector~\cite{Alves:2008zz,LHCb-DP-2014-002}, schematically shown in
figure~\ref{fig:detector} is a single-arm forward spectrometer covering the pseudorapidity range $2
< \eta < 5$, designed for the study of particles containing $b$ or $c$ quarks. The detector includes a
high-precision tracking system consisting of a silicon-strip vertex detector (VELO) surrounding the
$pp$ interaction region, a large-area silicon-strip detector (TT) located upstream of a dipole
magnet with a bending power of about 4\,Tm, and three stations (T1-T3), consisting of silicon-strip
detectors (IT) and straw drift tubes (OT) placed downstream of the magnet. The tracking system
provides a measurement of momentum, $p$, of charged particles with a relative uncertainty that
varies from 0.5\,\% at low momentum to 1.0\,\% at 200\,GeV.
The minimum distance of a track to a primary vertex (PV), the impact parameter (IP),
is measured with a resolution of $(15+29/p_{\rm T})\mu m$, where $p_{\rm T}$ is the component of the
momentum transverse to the beam, in\,GeV.\\
Different types of charged hadrons are distinguished
using information from two ring-imaging Cherenkov detectors (RICH1/2). Photons, electrons and
hadrons are identified by a calorimeter system consisting of scintillating-pad (SPD) and preshower
detectors (PS), an electromagnetic calorimeter (ECAL) and a hadronic calorimeter (HCAL). Muons are
identified by a system composed of alternating layers of iron and multiwire
proportional chambers (M1-M5).\\
The LHCb coordinate system is a right handed Cartesian system with the origin at the interaction
point. The $x$-axis is oriented horizontally towards the outside of the LHC ring, the $y$-axis is
pointing upwards with respect to the beam line and the $z$-axis is aligned with the beam direction.\\
The online event selection is performed by the trigger system, which is composed of three stages: a stage
implemented in hardware known as Level 0 (L0) and two stages implemented in software called High-
Level-Trigger (HLT1 and HLT2). The HLT1 performs a partial reconstruction of the candidates. In
this stage most requirements are inclusive, which means that the selection is applied only to subset of
the final state particles. A few exclusive algorithms are used to select specific decays at this
trigger stage. The HLT2 contains hundreds of inclusive and exclusive algorithms which are more
time-consuming and provide dedicated output for offline analyses.

\subsection{Track reconstruction}
\begin{figure}[h]\centering
  \includegraphics[width=0.7\textwidth]{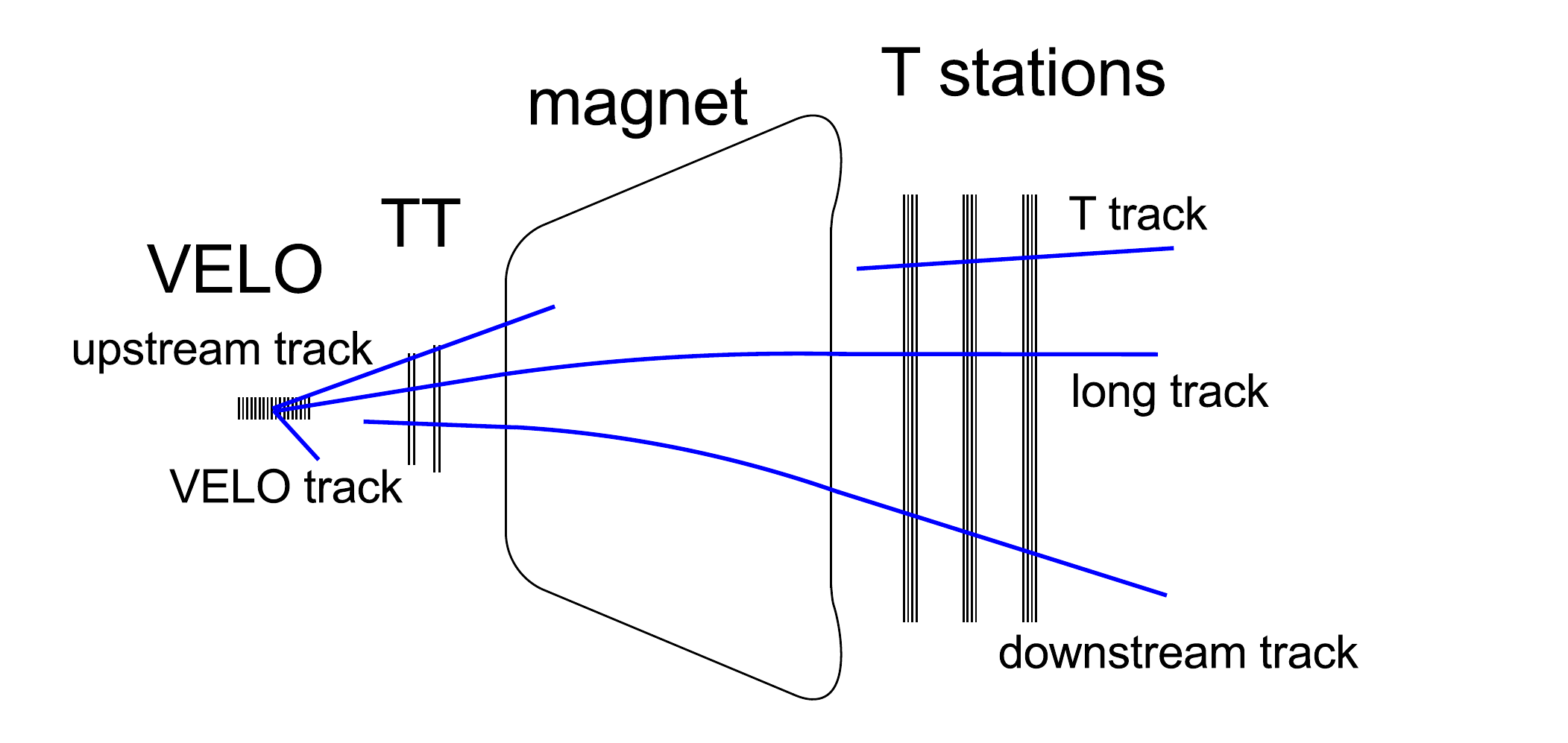}
  \caption{Track types in LHCb. Long tracks and downstream tracks are used for most physics
  analyses, the other types either serve as a component of another track type or are mainly used for
  detector studies.}
  \label{fig:tracktypes}
\end{figure}

Figure \ref{fig:tracktypes} shows an overview of the
different track types defined in the LHCb reconstruction: VELO tracks, which have hits in the
VELO; upstream tracks, which have hits in the VELO and TT; T tracks, which have hits in
the T stations; downstream tracks, which have hits in TT and the T stations; and long tracks, which
have hits in the VELO and the T stations. The latter tracks can additionally have hits in TT.
\\
Long tracks are the highest quality tracks comprising all available information from the trackers
and are therefore used in most physics analyses. Downstream tracks mainly play a
role in the reconstruction of daughters from long-lived particles which have decayed after the
VELO (usually weakly decaying strange hadrons, such as $\Lambda^0$ or $K_{\rm S}^0$).\\
Track reconstruction can be subdivided into a track finding/pattern recognition part and a track
fitting part which is done by a Kalman filter. The basic track finding algorithms, called VELO
tracking~\cite{Callot:1322644} and T seeding~\cite{Callot:1119095}, reconstruct VELO and T track
candidates which are used as seeds for upstream, long and downstream tracks. Long track candidates
are found by two dedicated algorithms. The first, called forward tracking~\cite{Callot:1033584},
starts with VELO or upstream tracks~\cite{Bowen:2105078} and searches for corresponding
hits in the T stations.
The second, called track matching~\cite{Needham:1020304,Needham:1060807}, uses both VELO and T
tracks as input and matches them in the magnet region.
Downstream tracks use T tracks as seed and searches for corresponding clusters in the
TT~\cite{Callot:1025827}. The outputs of all algorithms are merged,
eliminating candidates that were found twice.

\subsection{Reconstruction sequence in Run II and the upgrade}
\begin{figure}[h]
  \includegraphics[width=0.25\textwidth]{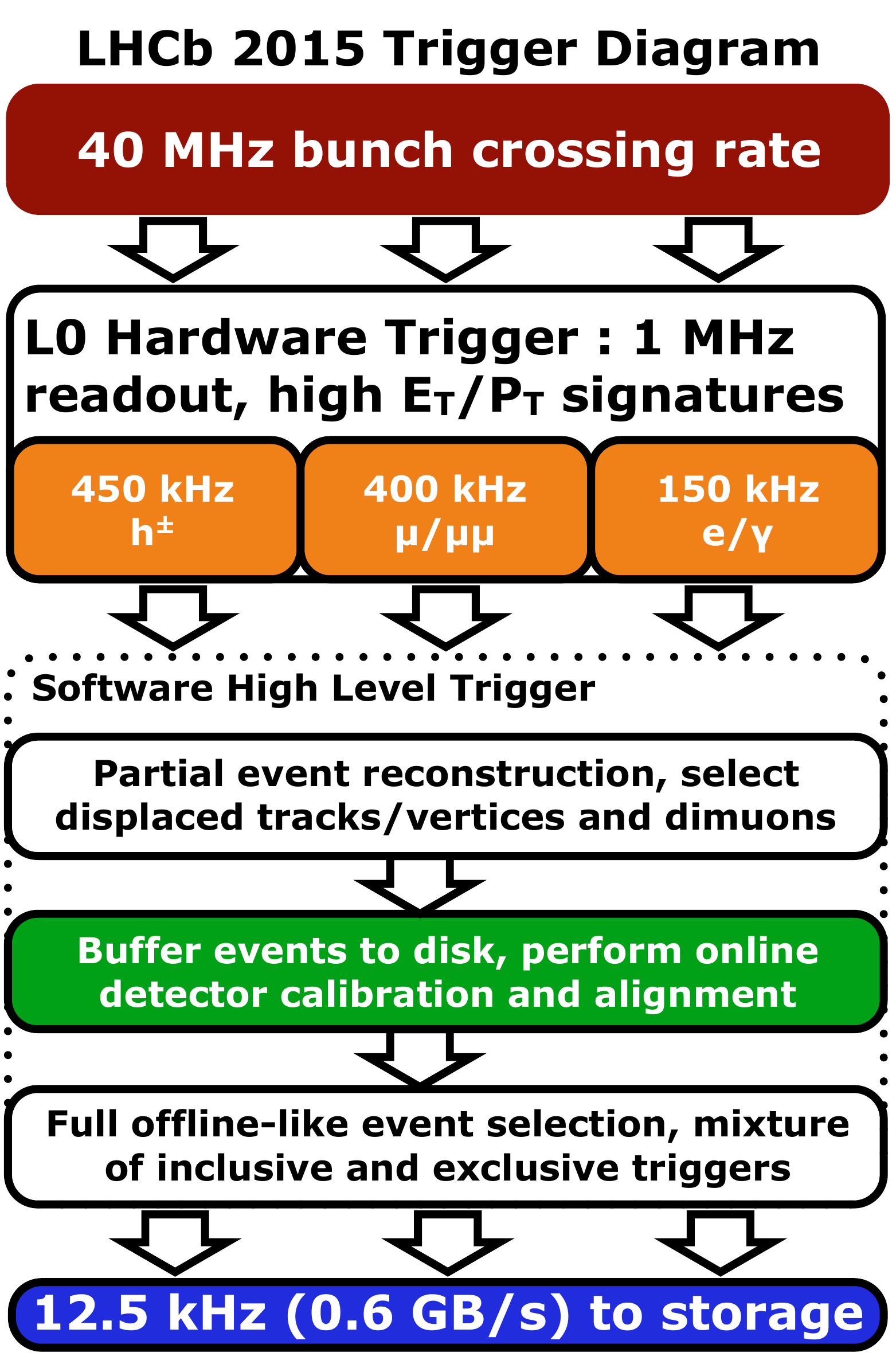}
  \raisebox{7mm}{\includegraphics[width=0.48\textwidth]{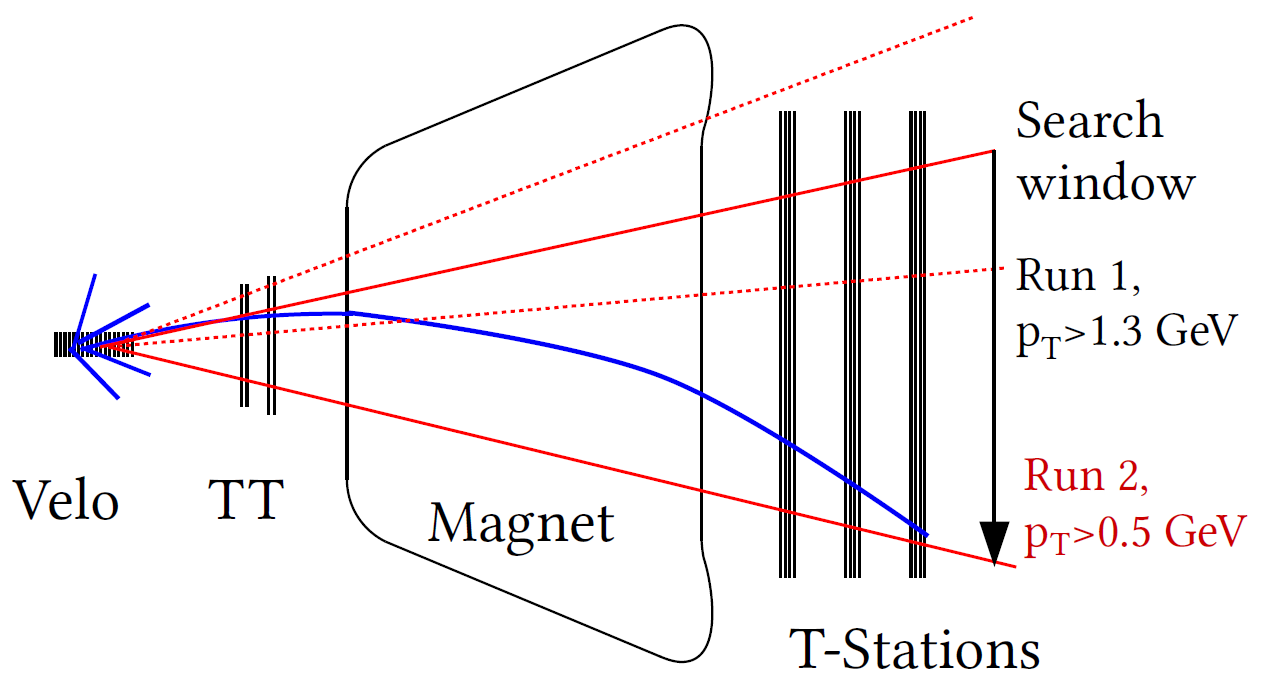}}
  \includegraphics[width=0.25\textwidth]{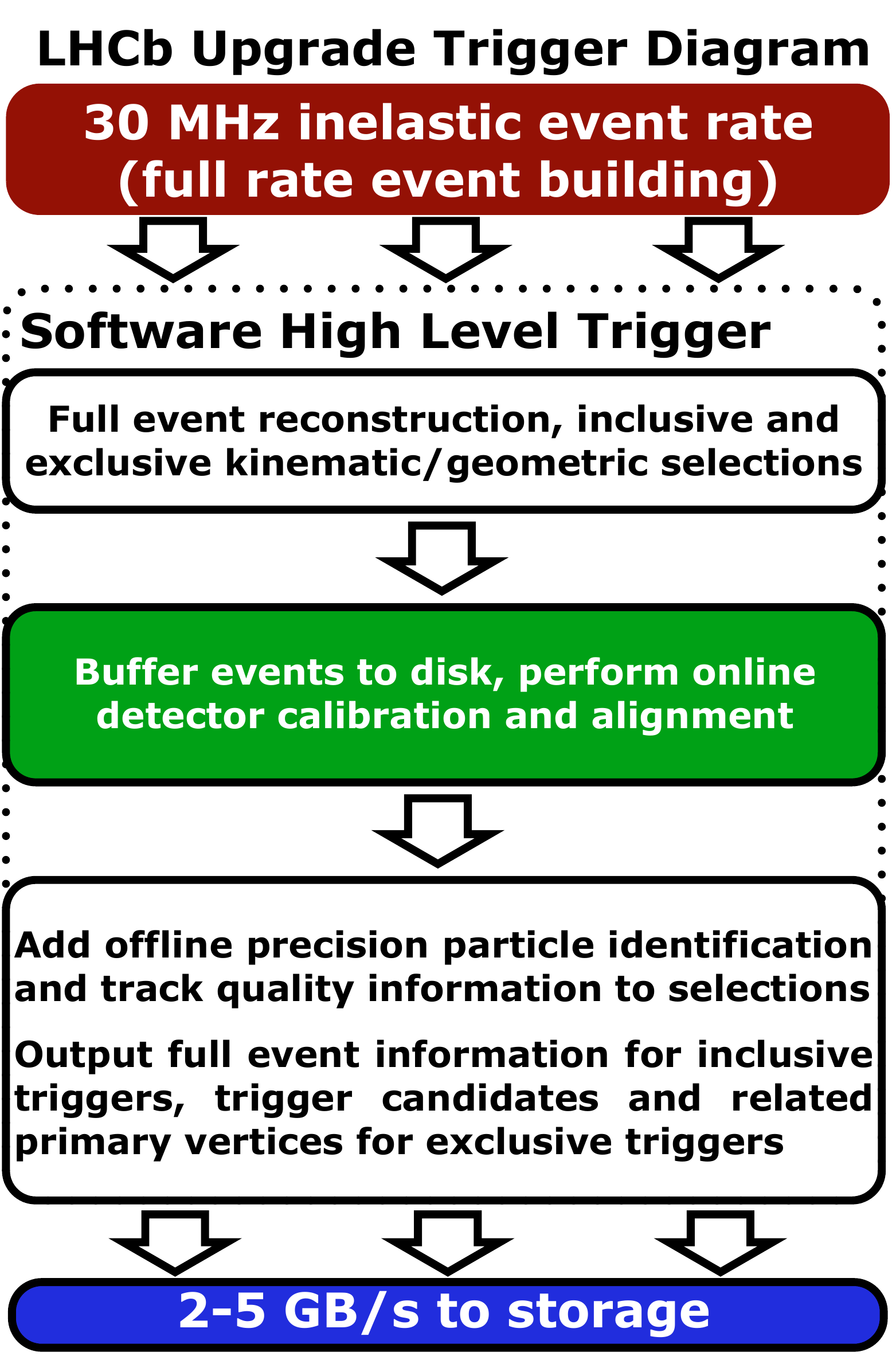}
  \caption{LHCb trigger scheme for Run II (left) and Run III (right). Long track
  reconstruction with the forward tracking algorithm in HLT1 comparing Run I and Run II (middle).}
  \label{fig:trg}
\end{figure}

The Run II trigger, schematically shown on the left of figure~\ref{fig:trg}, uses different track
reconstruction sequences in the fast (HLT1) and full (HLT2) stage of the software trigger. In HLT1,
all VELO tracks are reconstructed and fitted with a simplified Kalman filter, allowing for single
rescattering at the sensor planes, and then used to find the primary vertices. VELO track
trajectories are then extrapolated to the TT to reconstruct upstream tracks. In addition, the charge
of the track can be estimated due to the magnetic fringe field in the TT. The upstream track
candidates are then used as input to a fast version of the forward tracking algorithm, where only
long track candidates with $p_{\rm T} > 500$\,MeV are accepted. The found long track candidates are
fitted with a Kalman filter.\\
The timing and fake rate of the HLT1 track reconstruction sequence in Run~II profits from
requiring a minimal transverse momentum, so that the forward tracking has to process only half of
the tracks~\cite{Storaci:2017850}. Furthermore, by using the charge estimate of the upstream track
candidate, the search window for potential hits only covers the region of the detector in which the
particle will be deflected by the magnetic field as shown in the middle of figure~\ref{fig:trg}.
Another positive effect of using upstream tracks in the fast reconstruction sequence is that fake
VELO tracks can be vetoed, since they are less likely to be confirmed with clusters in TT.\\
The HLT2 sequence of the forward tracking runs - with looser requirements - on VELO tracks that
could not be promoted to long tracks in the HLT1 sequence. Details are discussed in section
\ref{sec:mlft}. In addition, standalone T tracks are matched with VELO
tracks, to form long tracks, and with TT clusters, to form downstream tracks. All track candidates
are fitted with a Kalman filter and clones from different algorithms are removed.
\\
With the LHCb upgrade programme for Run~III data taking, all three tracking detectors will be
replaced. The current silicon strip VELO will be replaced by a pixel-based
solution~\cite{LHCb-TDR-013}; the TT by a silicon strip detector with higher granularity, called
Upstream Tracker (UT); and the T stations by a Scintillating Fiber Tracker (SciFi), read out by
silicon photomultipliers~\cite{LHCb-TDR-015}.\\
In Run III, the LHCb experiment will take data at an instantaneous luminosity of
2$\cdot$10$^{33}$\,cm$^{-2}$\,s$^{-1}$ - five times as high as in Run~II. At this luminosity, the
rate of potentially interesting physics would be too high to have a reasonable compromise of event
reduction and efficiency with the simple selections of L0 hardware trigger. It was therefore decided
to move to a trigger-less readout system and a full software trigger as shown on the right of
figure~\ref{fig:trg}. This puts strong constraints on the execution time of the tracking sequence in
the fast trigger stage, which - at a higher occupancy - is expected to be reduced by a
factor $\sim$2.5 compared to the current sequence. It will thus require massive use of
cost-effective parallel computing techniques as well as finding an optimal working point in the
trade-off between timing and track reconstruction efficiency at a low fake rate.

\section{Parallelism}\label{sec:par}
Although the reconstruction sequences became faster with the changes described in the previous
section, further timing improvements were required. Data level parallelism in modern multi-core CPUs
in the form of Single Instruction Multiple Data (SIMD) instructions are used to speed up time consuming
parallelizable parts of the reconstruction code. These are purely mathematical
computations requiring a significant amount of CPU cycles. However, most of the time consuming parts
of the track finding algorithms are not parallelizable, since they are contained in bodies of
conditional statements which are predicate to whether a computation is executed or not.
\subsection{SIMD in forward tracking}
Each of the tracking stations has four detection layers in an ($x$-$u$-$v$-$x$)
arrangement with vertically oriented modules in the first and the last layer and rotated modules by
a stereo angle of -5$^\circ$ and +5$^\circ$ in the second and the third layer, respectively.
In the forward tracking algorithm, trajectories defined by a seed track (either VELO or upstream
tracks) and a hit in the $x$-layers of the T stations, are projected through the magnetic field
into a plane parallel to the tracking stations at a given $z$ position ("Hough plane"). For an ideal
magnet, the track trajectory outside the magnetic field could be described by two straight lines, whose
extrapolations intersect in the middle of the magnet, as shown in figure~\ref{fig:maghough}. But
magnetic fringe fields well outside the magnet volume, reaching at least up to the first T station,
force the use of an empirically found cubical parametrization of the projection trajectory. This
projection of seed tracks is done for every hit in the $x$ layers of the tracking stations
independently and is therefore an optimal use case for SIMD instructions.
A timing improvement of 40\,\% has been reached in this part of the code. This is close to
the theoretical maximum of 50\,\%, using \texttt{double}s in the lowest common instruction set
extension on the HLT computing farms - SSE2.
Some percent in timing are lost by arranging the data into SIMD-usable form.

\subsection{SIMD in track fitting}
\begin{figure}
\centering
\begin{minipage}{.45\textwidth}
  \centering
  \includegraphics[width=\linewidth]{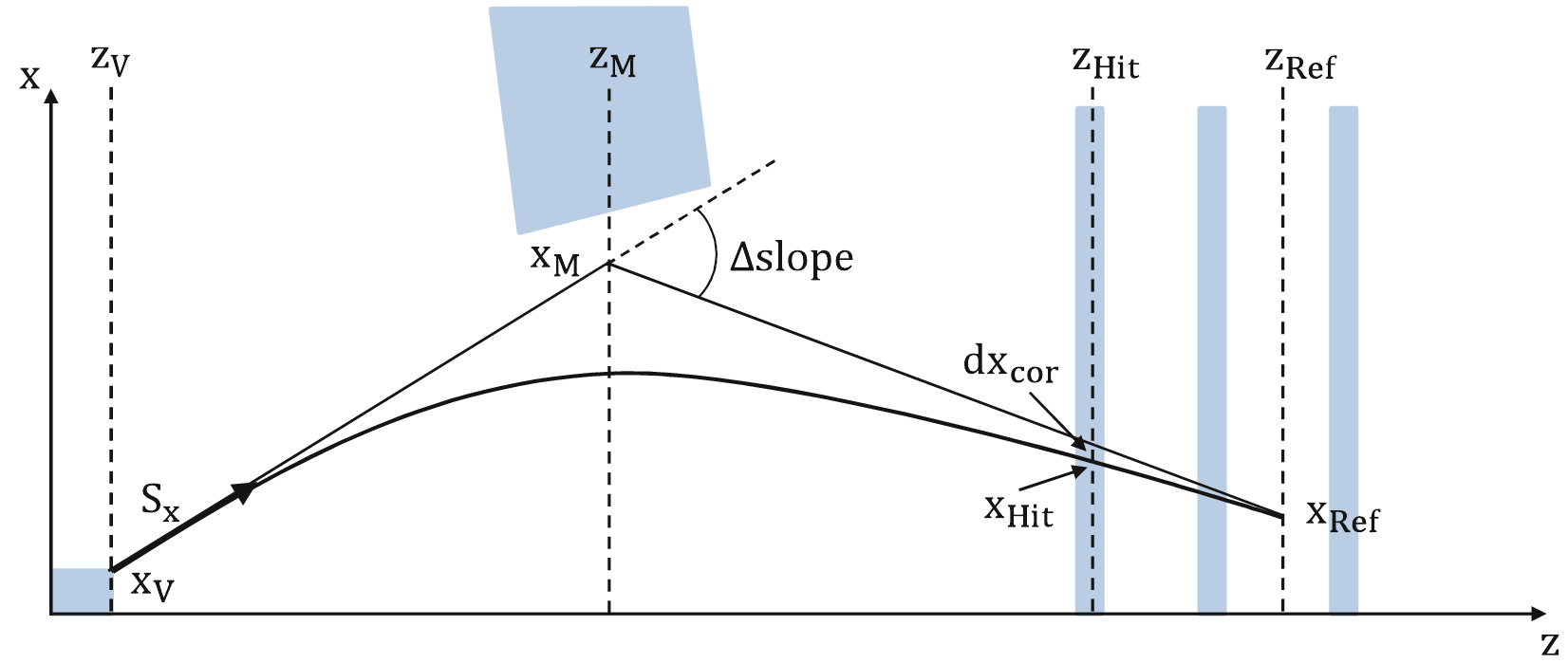}
  \caption{\label{fig:maghough}Sketch of seed track projection through the magnetic field. The true track trajectory is
  shown as curved line, whereas the simplified projection is given by straight lines intersecting
  in the middle of the magnet.}
\end{minipage}\hspace{2pc}%
\begin{minipage}{.45\textwidth}
  \includegraphics[width=\linewidth]{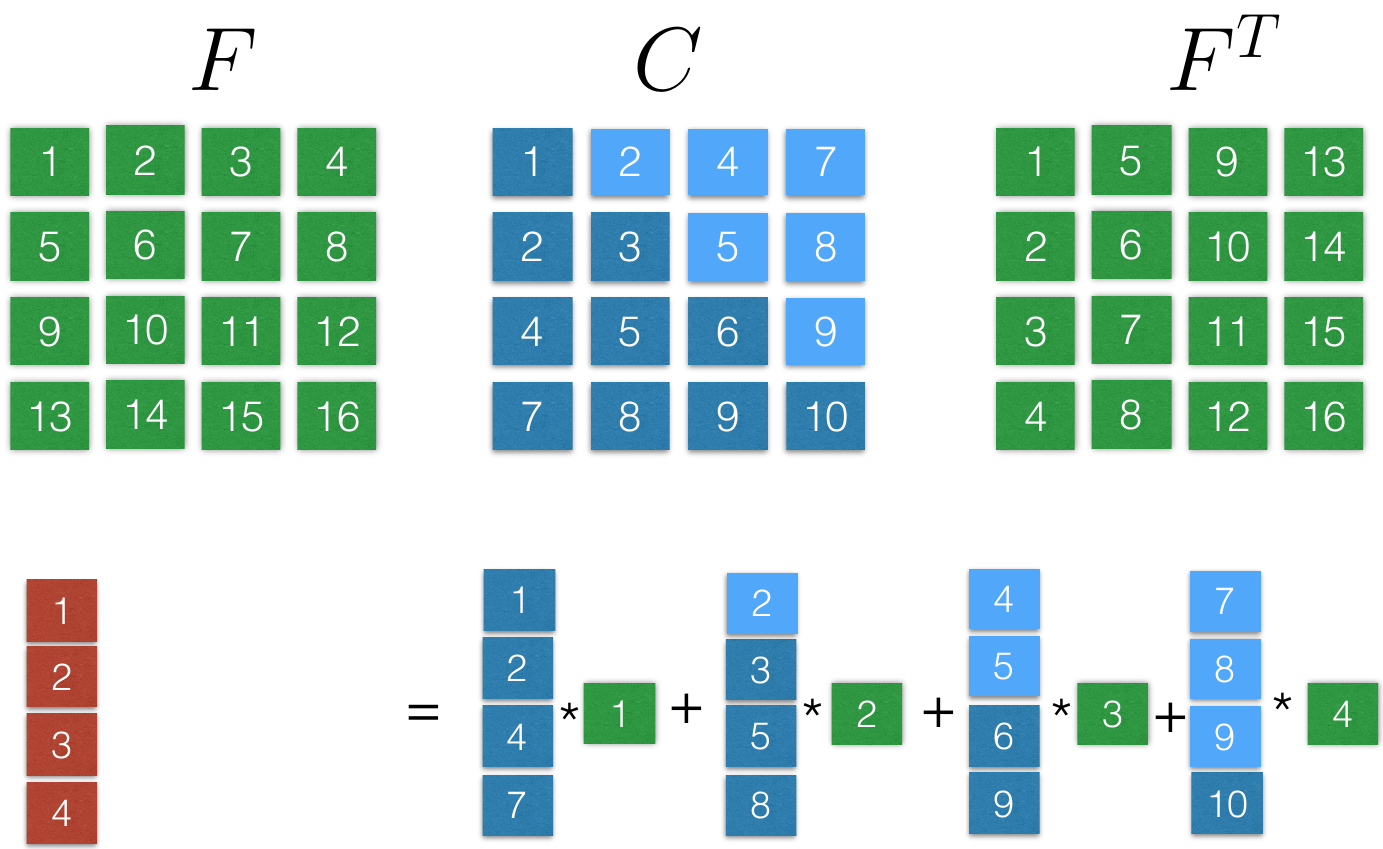}
  \caption{\label{fig:fcft}Illustration of first parallelizable step of the transform $F\cdot C\cdot F^{\rm T}$.}
\end{minipage}%
\end{figure}

Track fitting is done with a Kalman filter to get the best track-parameter estimates. The fitting
stage is the largest timing contributor in the HLT1 reconstruction sequence; mainly due to expensive
calculus such as solving differential equations for the propagation through the magnetic field with
the Runge-Kutta method or 5$\times$5 matrix operation needed by the Kalman filter. SIMD instructions
are used for transportation of the covariance matrix from state $k$ to state $k+1$. This operation
requires matrix multiplications of the form $F\cdot C\cdot F^{\rm T}$, with a transport matrix $F$
and a symmetric covariance matrix $C$. The first parallelizable step of multiplying $C$ with the
first column of $F^{\rm T}$ for 4$\times$4 matrices is shown in figure~\ref{fig:fcft} and amounts to
16 multiplications and 3 additions. This reduces to 7 operations using the Advanced Vector
Extensions (AVX) instruction set, which allows to process 4 \texttt{double} values in parallel. In
the case of 5$\times$5 matrices, reordering the multiplications such that they are vectorized
reduces the time consumption by about a factor of 2, while AVX would lead to a factor 5.

\section{Machine learning}\label{sec:ml}
\subsection{Fake track rejection}
\begin{figure}[h]
  \includegraphics[width=0.48\textwidth]{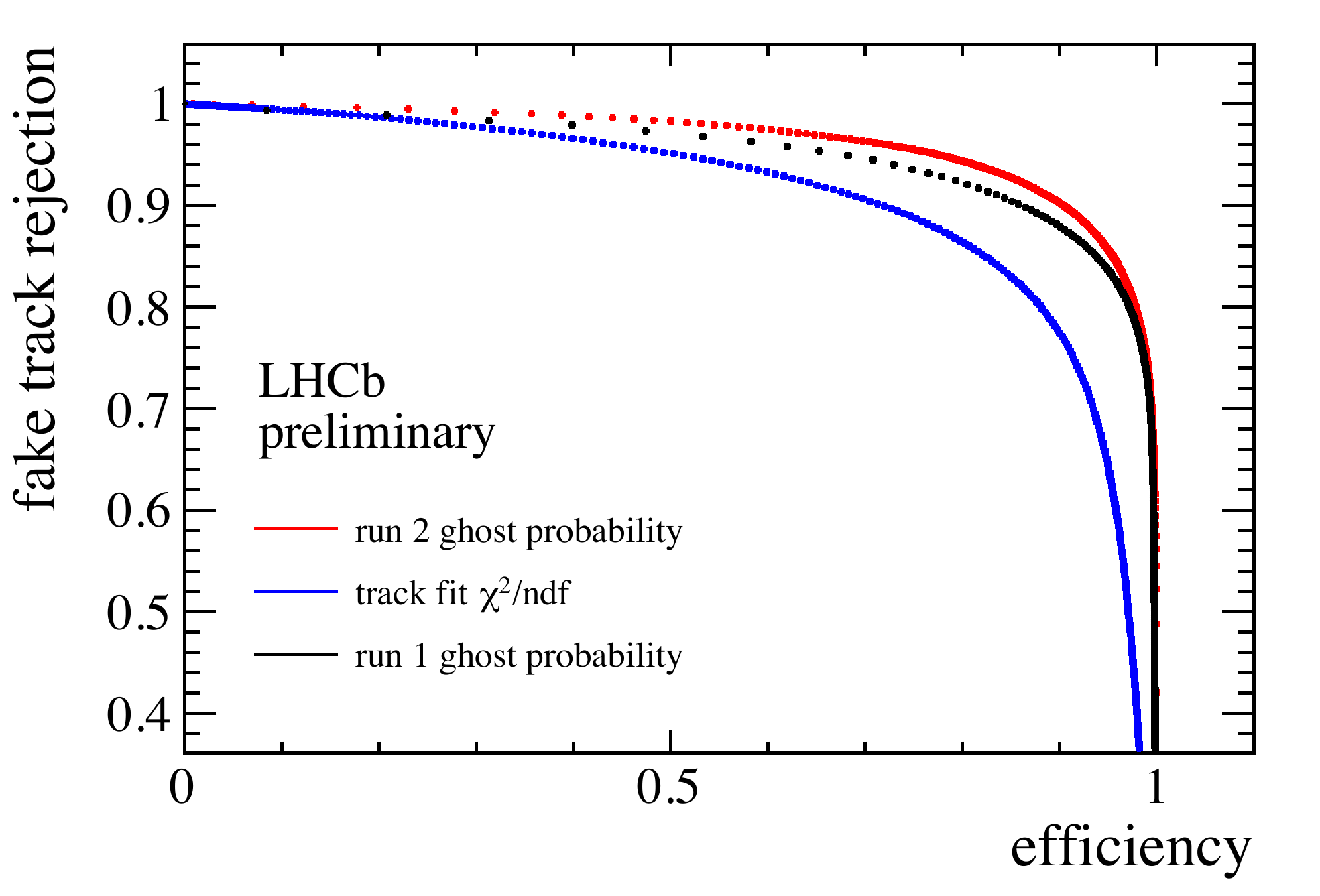}
  \includegraphics[width=0.48\textwidth]{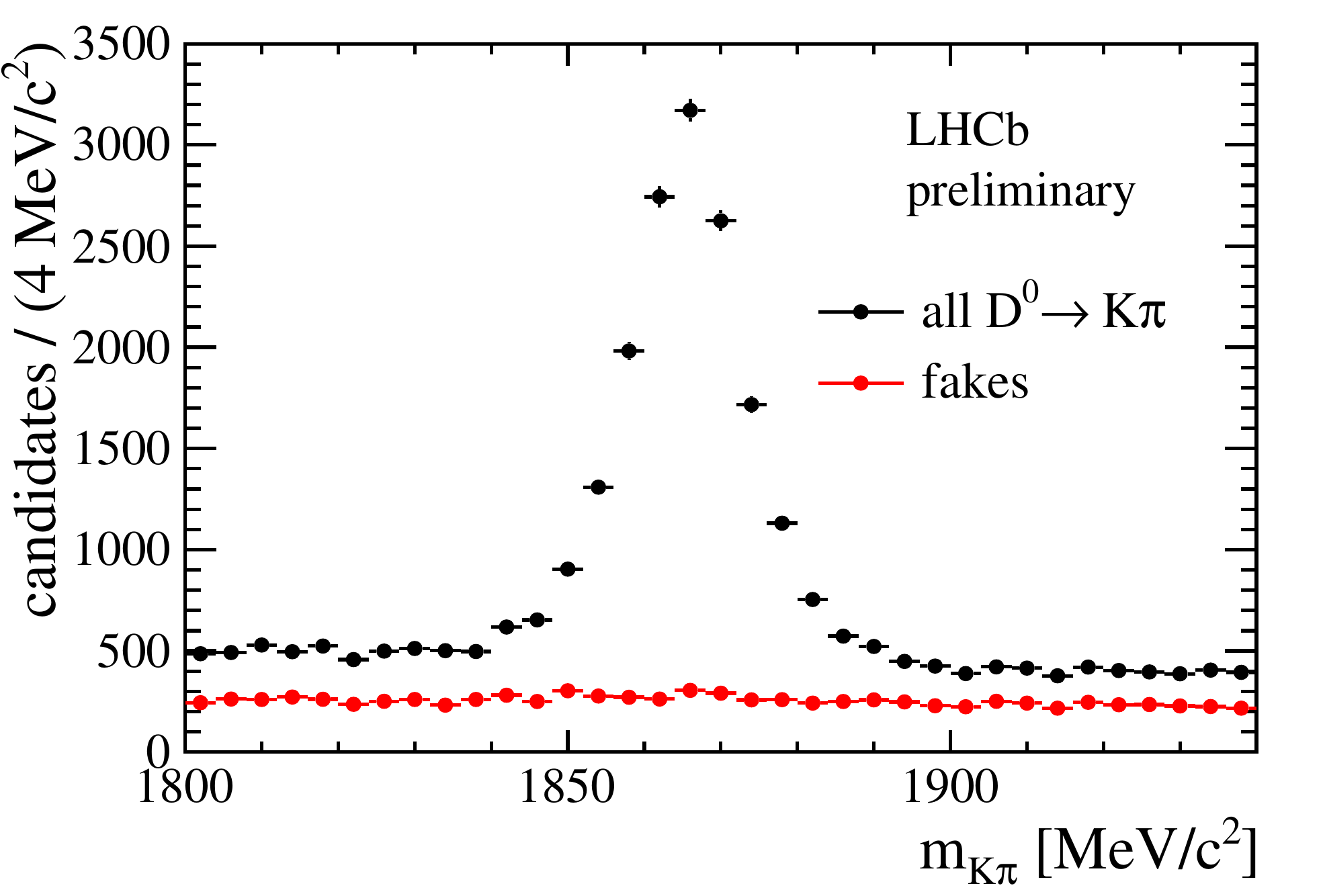}
  \caption{ROC curves for long tracks for the Run~II ghost probability, the track
  $\chi^2$/nDoF and the Run~I ghost probability obtained on a Run~II dataset (left). At the optimal
  working point the fake track rate could be reduced from 22\,\% to 14\,\%, when comparing a cut on the track
  $\chi^2$/nDoF and the ghost probability.\\
  $D^0 \to K\pi$ candidates from HLT2 output without cut on the ghost probability and events
  rejected by a cut on the ghost probability of 30\% (right).}
  \label{fig:GP}
\end{figure}

Fake long tracks can originate from falsely reconstructed track segments in the VELO or the
T stations, from a mismatch of VELO and T station segments or from hadronic interaction of
particles with the detector. During reconstruction, most of the fake long tracks originate from
hadronic interactions, followed by fake track segments in the T stations and mismatched segments,
while fake reconstruction of VELO segments occur at lower rate.
Mismatched track segments are due to the long lever arm between the tracking stations up- and
downstream of the magnet and remain to be the most abundant category after fake track rejection.\\
To discriminate between good and fake tracks, a fast artificial neural network classifier in form of
TMVA's~\cite{Hocker:2007ht} MultiLayerPerceptron (MLP) is used in both stages of the software
trigger after track fitting. While the "ghost probability" - the name of the MLP's response - was
an offline quantity during Run~I, a huge speed-up by a factor $\sim$90 allowed for its processing
in HLT2 in 2015. A cut on the track $\chi^2$/nDoF of the track fit was used to reject ghost tracks
in Run~I and in HLT1 in 2015. Since 2016 the fake track rejection also runs in the HLT1
reconstruction sequence, leading to a significant speed-up of subsequent algorithms.\\
Main advances in mentioned speed-up come from choosing $1/\sqrt{1+x^2}$, rather than $\tanh(x)$ as
activation function, amounting to a factor 50 in terms of CPU-cycles; but also from choosing input
variables which themselves are already available or do not require much additional computing time; and
finally from manual optimization of the automatically generated class file for the TMVA reader~\cite{paulgh}.
Using AVX instructions, given the $1/\sqrt{1+x^2}$ activation function, the number of CPU cycles
could potentially be reduced by more than a factor of 2 in future applications.\\
The MLP uses 21 input variables and one hidden layer with 26 nodes. No performance gain from a deeper
hidden layer structure, more layers, or human assisted learning was observed.
To obtain a physical interpretation to the response, a probability integral transform - also
referred to as "flattening" or "rarity transformation" - is obtained as a linear spline fit
to the cumulative network response for fake tracks in simulated events.
The performance of the artificial neural network is illustrated in figure~\ref{fig:GP}.
\clearpage
\subsection{Machine learning in the forward tracking}\label{sec:mlft}
The forward tracking algorithm has undergone a major revision for 2016 data taking. Apart from
re-tuning its parameters to adapt to Run~II conditions, two artificial neural networks have been
implemented to increase reconstruction efficiency and reject fake track candidates in the early
stages of the reconstruction.\\
The forward tracking uses upstream tracks as seed tracks in HLT1 and VELO tracks as seeds in HLT2.
A search window in the T stations is computed based on the seed track information. All T station hits in the
$x$-layers are projected into a reference plane where - in form of a cluster search - $x$-track
candidates are built. A $\chi^2$ fit to these candidates is performed and hit-outliers are removed. After that,
hits in the $u$ and $v$ modules are added and a similar $\chi^2$ fit is performed, where outliers
are removed. Track candidates that pass certain quality criteria are stored in a container
which is an input to the Kalman filter. In HLT1, $x$-track candidates are only build if the
corresponding cluster had at least one hit in 5 of 6 T station $x$-layers. In HLT2 the clustering
runs with the same conditions, but if no long track candidate could be built from a seed track, a
recovery loop (RL) is run where also $x$-track candidates are taken into account, which come from
clusters with at least one hit in 4 T station $x$-layers.\\
One of the two neural networks is used to reject bad track candidates just before the track is
stored for the Kalman filter. This network is evaluated in both reconstruction sequences of the
trigger, using a slightly looser response cut in HLT2. The other network response is used to reject
bad track candidates in the recovery loop of the HLT2 sequence for $x$-track candidates with only
one hit in four different $x$-layers to reduce the large combinatorial background.\\
The artificial neural networks are again TMVA's MLPs which have been trained to optimize fake track
rejection at a given efficiency of 99 or 97\,\%. It was found that the classification performance
improved with a deep hidden layer structure and larger number of parameters, even though some of
them highly correlated. The MLP in the recovery loop has 9 input parameters and 2 hidden layers
with 16 and 10 nodes, whereas the MLP for the final candidate selection has 16 input parameters and
3 hidden layers with 17, 9 and 5 nodes. The Rectified Linear Unit (ReLU) ($\max(0,x)$) was chosen as
activation function to ensure fast computation of the network response.\\
The performance of both neural networks was evaluated on Monte Carlo and validated with minimum bias
data. Slight differences were found and the parameters of the forward tracking algorithm were
adjusted accordingly. The performance results of the so found set of parameters is summarized in
table~\ref{tab:one}. The neural networks contribute 0.5\,\% and 2\,\% to the execution time of the
forward algorithm. Even though the forward tracking became slightly slower, the time consumption of
the Kalman filter, and eventually the whole reconstruction sequence, was reduced due to the removal
of fake tracks.
\begin{table}[h]
\caption{\label{tab:one}Overview over the most important measures for the improved forward tracking.
Timing and fake rates are given relative (= $X_{new}/X_{ref}-1$) to the previous (2015) forward tracking.
The changes in efficiency are absolute changes. $\varepsilon$ long and $\varepsilon$ long from B
(i.e. from a hadron containing a $b$- or $\bar{b}$-quark) were extracted after all reconstruction
steps, while the HLT1 efficiency has to be evaluated at an intermediate step.
The absolute reference fake rate at this stage of the reconstruction was 33\,\% and 5.9\,\% in HLT1.
The absolute efficiencies given here are all $\sim$90\,\%.
Figures without recovery loop are given as optional setting.}
\begin{center}
\lineup
\vspace*{2mm}
\begin{tabular}{lrr}
  \toprule
  MC performance            & \multicolumn{2}{c}{ $\nu=$1.6 } \\
  2016 w.r.t. 2015          & \multicolumn{1}{c}{with RL} & \multicolumn{1}{c}{without RL} \\\midrule[\heavyrulewidth]
  timing HLT1               & \multicolumn{2}{c}{$\pm\,$0.0 \%}\\
  timing HLT2               & $+\,$4.0 \% & $-\,$38.0 \% \\\midrule
  fake rate                 & $-\,$26.9 \% &  $-\,$35.1 \%\\
  fake rate HLT1            & \multicolumn{2}{c}{$-\,$14.1 \%} \\\midrule
  $\varepsilon$ long        & $+\,$0.5 \% & $+\,$0.1 \%\\
  $\varepsilon$ long from B & $+\,$0.2 \% & $-\,$0.2  \% \\\midrule
  $\varepsilon_{\mathrm{HLT1}}$ \begin{tiny}long from B $p>$3,$p_T>$0.5 GeV \end{tiny} &
  \multicolumn{2}{c}{$+\,$0.1 \%}\\\bottomrule
\end{tabular}
\end{center}
\end{table}

\section{Conclusion}\label{sec:con}
The LHCb experiment moved to a new trigger strategy with a real-time reconstruction, alignment and
calibration for Run~II data taking. To maximize the output of interesting events from the software
trigger, the execution time of the track reconstruction sequences was decreased by an overall factor
of two. This crucial speed-up was achieved with the help of SIMD instructions to increase the speed
of the code and machine learning to efficiently remove fake tracks already in the early stages of
the reconstruction. Ever faster algorithms exploiting parallelism and machine learning will be
needed for Run~III data taking, where LHCb will move to a trigger-less readout system and a full
software trigger.
\vspace*{1.cm}
\section*{References}
\bibliographystyle{iopart-num}
\bibliography{main,LHCb-PAPER,LHCb-CONF,LHCb-DP,LHCb-TDR}

\end{document}